\documentclass[aps,10pt,twocolumn,prl,floats,floatfix,superscriptaddress]{revtex4-1}
\usepackage{amsmath}
\usepackage{amssymb}
\usepackage{braket}
\usepackage{bbold}
\usepackage{pifont}
\usepackage{enumitem}
\usepackage{verbatim}
\usepackage{listings}
\usepackage{graphicx}
\usepackage[usenames,dvipsnames,svgnames,table]{xcolor}
\usepackage[dvips]{epsfig}
\usepackage[T1]{fontenc}
\usepackage{shadow}
\usepackage{varioref}
\usepackage{epsfig}
\usepackage{array}
\usepackage{rotating}

\newcommand{\beqn}{\begin{eqnarray}}
\newcommand{\eeqn}{\end{eqnarray}}
\newcommand{\dd}{\mathrm{d}}
\newcommand{\nn}{\nonumber}
\newcommand{\Tr}{\mathrm{Tr}}

\newcommand{\gmn}{g_{\mu\nu}}

\newcommand{\fmn}{f_{\mu\nu}}

\newcommand{\hmn}{h_{\mu\nu}}

\begin{document}

\title{Interactions of multiple spin-2 fields beyond pairwise couplings}

\author{S.F.~Hassan}
\affiliation{Department of Physics \& The Oskar Klein Centre, Stockholm
  University, AlbaNova University Centre, SE-106 91 Stockholm, Sweden}

\author{Angnis~Schmidt-May}
\affiliation{Max-Planck-Institut f\"ur Physik
  (Werner-Heisenberg-Institut), F\"ohringer Ring 6, 80805 Munich,
  Germany}


\begin{abstract} 



Thus far, all known ghost-free interactions of multiple spin-2 fields
have involved at most pairwise couplings of the fields, which are
direct generalizations of bimetric interactions. Here we present a
class of spin-2 theories with genuine multi-field interactions, and         
explicitly demonstrate the absence of ghost instabilities. The
construction involves integrating out a nondynamical field in a theory
of spin-2 fields with only pairwise ghost-free interactions. The new
multivierbein interactions generated are not always expressible in
terms of the associated metrics.
\end{abstract}

\maketitle



\section{Introduction }


Interacting theories for multiple fields with spin 0, $1/2$ and 1 are
well understood and realized in Nature via the Standard Model of
Particle Physics, where the multiplets and their mixings are crucial
for the viability of the theory. In contrast, General Relativity is
the simplest possible theory of a single spin-2 field.  It is a
fundamental question whether, in analogy with lower spins, consistent
theories of multiple spin-2 fields could exist. Such theories could
have profound implications for the understanding of the gravitational
force beyond General Relativity, but have not been easy to construct.

In a covariant set up, spin-2 fields have more components than
physically needed, and generic theories do not have enough symmetries
and constraints to remove the unphysical components. Some of these, if
not eliminated, give rise to ghost instabilities, an example being the
Boulware-Deser ghost \cite{Boulware:1973my} of a massive spin-2
field. A few years ago the ghost-free theory of two interacting spin-2
fields was found \cite{Hassan:2011zd}, which also fulfills some other
important consistency criteria
\cite{Hassan:2017ugh,Kocic:2018yvr}. This generalized previous work on
a single \textit{massive} spin-2 field in a fixed background
\cite{deRham:2010kj,Hassan:2011vm,Hassan:2011hr, Hassan:2011tf}. The
model is formulated in terms of two symmetric rank-two tensors (or
``metrics") $\gmn$ and $\fmn$ interacting through a specific potential
$V_\mathrm{bi}(g,f)$, hence the name ``ghost-free bimetric
theory''. For a recent review see \cite{Schmidt-May:2015vnx}.

Theories for more than two spin-2 fields are also strongly restricted
by the absence of ghosts.  From the analysis of \cite{Hassan:2011zd},
it is easy to see that certain ghost-free theories can be constructed
as straightforward extensions of bimetric theory, by simply adding
copies of the bimetric potentials $V_\mathrm{bi}(g,f)$ for pairs of
metrics but without forming loops. An example for four metrics
$g^I_{\mu\nu}$ would be $V_\mathrm{bi}(g^1,g^2) +
V_\mathrm{bi}(g^1,g^3)+ V_\mathrm{bi}(g^3,g^4)$. So far, these
pairwise couplings have been the only known ghost-free interactions
of multiple spin-2 fields.

An important class of multi spin-2 theories was constructed in
\cite{Hinterbichler:2012cn}, using antisymmetrized products of the 
vierbein fields. These appeared to be ghost-free, however, a more
detailed analysis in \cite{deRham:2015cha} revealed that generically
such multivierbein interactions contained ghosts. It was argued 
that the ghost-free subset consisted only of models where the
vierbeins could be traded off for the metrics by virtue of a
\textit{vierbein symmetrization condition}, exactly as in bimetric
theory. The only known models of this type are the pairwise
interactions described above. Hence the question is if genuine multiple
spin-2 interactions beyond the pairwise ones, and beyond the class
conjectured in \cite{deRham:2015cha}, exist. Here we show that this
is indeed the case.

\paragraph{Summary of results.}
In this work we derive a class of ghost-free interactions for multiple
spin-2 fields by integrating out a non-dynamical field in a theory
with ghost-free bimetric interactions.  The result is an interaction
term for $\mathcal{N}$ vierbeine $(e_I)^A_{~\mu}$ of the form,
\begin{align}\label{vbint}
  S_\mathrm{multi}=-M^4\int\dd^4x~\det\Big(\sum_{I=1}^\mathcal{N}
  \beta^{I}e_I\Big)\,,
\end{align} 
involving a mass scale $M$ and arbitrary dimensionless coefficients
$\beta^I$, $I=1,\hdots,\mathcal{N}$.  The kinetic terms of the
vierbein fields have the standard Einstein-Hilbert form. For
restricted vierbein configurations the multi-vierbein vertex can be
expressed in terms of metrics, but this is not always possible. The
interactions can involve up to four different vierbeine in each term
and are therefore more general than the pairwise couplings known so
far.


\section{Generating new interactions}


The starting point is a theory for $(\mathcal{N}+1)$ vierbeine,
$u_{~\mu}^A$ and $(e_I)_{~\mu}^A$, $I=1,\hdots \mathcal{N}$,  
with ghost-free bimetric interactions. Let us denote the
corresponding metrics by $f^0_{\mu\nu}=u_{~\mu}^A\eta_{AB}u_\nu^{~B}$
and $\fmn^I=(e_I)_{~\mu}^A\eta_{AB}(e_I)_{~\nu}^B$. The action has the 
following structure,  
\begin{align}\label{np1}
S[u, e_I]= \sum_{J=0}^\mathcal{N} S_\mathrm{EH}[f^J]
+S_\mathrm{int}[u,e_I]\,.
\end{align}
It includes the Einstein-Hilbert kinetic terms,
\begin{align}\label{EH}
S_\mathrm{EH}[f^J]=m_J^2\int\dd^4x~\sqrt{f^J}~R(f^J)\,,
\end{align}
where, $m_J$ are the $(\mathcal{N}+1)$ Planck masses. $S_\mathrm{int}$   
contains the simplest ghost-free bimetric interactions between
$u_{~\mu}^A$ and each one of the $(e_I)_{~\mu}^A$,
\begin{align}
\label{intact}
&S_\mathrm{int}[u,e_I] = \nn\\
&\qquad -2m^4\int\dd^4x\,\det u\,\Big(\beta_0+\sum_{I=1}^\mathcal{N} 
\beta^I\Tr\big(u^{-1}e_I\big)\Big)\,.
\end{align}
In addition to the traces $\Tr(u^{-1}e_I)=u^{\mu}_{~A}(e_I)_{~\mu}^A$,
we could also include the remaining ghost-free bimetric interactions,
but have chosen not to do so. For brevity, we set $\beta^I=1$ for
$I=1,\hdots,\mathcal{N}$, by scaling $e_I\rightarrow e_I/\beta^I$,
and redefining the Planck masses $m_I$ in (\ref{EH}) accordingly.


\paragraph{Lorentz constraints.}
Each vierbein contains $6$ Lorentz parameters that drop out of the
corresponding metric and hence appear in the action \eqref{np1} only
through the potential terms in $S_\mathrm{int}$. Since these are
nondynamical, their equations of motion are constraints. Specifically,
these are the antisymmetric parts of the equations of motion for
$(e_I)_{~\mu}^A$ \cite{Zumino:1970tu,Hinterbichler:2012cn}, precisely
6 equations per vierbein,
\begin{align}\label{constlor}
\frac{\delta S_\mathrm{int}}{\delta
  (e_I)_{~\mu}^{A}}\,\eta^{AB}\,(e_I)^\nu_{~B}-\frac{\delta
  S_\mathrm{int}} {\delta
  (e_I)_{~\nu}^{A}}\,\eta^{AB}\,(e_I)^\mu_{~B}=0\,.
\end{align}
The corresponding equation for $u_{~\mu}^A$ is a linear combination of
(\ref{constlor}), due to the overall Lorentz invariance of the action.
For the potential in (\ref{intact}), the Lorentz constraints 
uniquely imply the following symmetrization conditions,
\begin{align}
\label{symc}
u_{~\mu}^A\eta_{AB}(e_I)_{~\nu}^B=u_{~\nu}^A\eta_{AB}(e_I)_{~\mu}^{B}\,,
\quad I=1,\hdots \mathcal{N}\,.
\end{align}
These allow us to express the potential in terms of metrics by
replacing $u^{-1}e_I=\sqrt{f_0^{-1}f_I}$ in addition to $(\det
u)=\sqrt{|\det f_0|}$. It is then straightforward to verify that the
arguments for the absence of ghost in bimetric theory
\cite{Hassan:2011zd} extend to this case.


\paragraph{Multi-spin-2 action.}
Let us take the limit $m_0\rightarrow 0$ to freeze out the dynamics of
the vierbein $u_{~\mu}^A$.  We can then eliminate $u_{~\mu}^A$
algebraically using its equation of motion to obtain an action for the
remaining $\mathcal{N}$ vierbeine. Indeed, varying the action
(\ref{np1}) with respect to the inverse vierbein $u^\mu_{~A}$ gives an
equation with the unique solution
\begin{align}\label{usol}
u_{~\mu}^A=-\frac{3}{\beta_0}\sum_{I=1}^\mathcal{N} (e_I)_{~\mu}^A\,.
\end{align}
Using this to eliminate $u^A_{~\mu}$ in \eqref{intact}, gives the new 
interactions for the remaining $\mathcal{N}$ vierbeine,    
\begin{align}\label{eff3}
S_\mathrm{int}[e_I]
= - M^4\int\dd^4x~\det \left(e_1+ e_2+ \hdots+ e_\mathcal{N}\right)\,,
\end{align}
with $M^4\equiv-54 m^4\beta_0^{-3}$. Although these are not linear
combinations of the pairwise bimetric potentials, we nevertheless
expect 
them to be
ghost-free since they were derived from a ghost-free
set up. We will explicitly confirm this below.

The $\mathcal{N}$ symmetrization constraints (\ref{symc}), with
$u^A_{~\mu}$ given by (\ref{usol}), provide a set of constraint
equations for the new multi-spin-2 theory. Equivalently, one can
derive these from the multi-spin-2 action \eqref{eff3} using
(\ref{constlor}). In matrix notation these new Lorentz constraints
are,
\begin{align}\label{consteff}
\sum_{I=1}^\mathcal{N} e_J^\mathrm{T}\eta e_I
=\sum_{I=1}^\mathcal{N} e_I^\mathrm{T}\eta e_J\,,
\quad J=1,\hdots,\mathcal{N}\,.
\end{align} 
Note that if we sum (\ref{consteff}) over $J$, the resulting equation
is identically satisfied. Hence there are $\mathcal{N}-1$ independent
matrix relations among the antisymmetric parts of the combinations
$(e_J)_{~\mu}^{A}\eta_{AB}(e_I)_{~\nu}^B$. These can be used to
eliminate $6(\mathcal{N}-1)$ non-dynamical components of the
vierbeine. Another 6 components are removed by the overall Lorentz
invariance of the action. This leaves us with
$16\mathcal{N}-6\mathcal{N}=10\mathcal{N}$ independent components,
which is the same as the number of components in $\mathcal{N}$
symmetric rank-2 tensor fields.


\section{Direct ghost proof}
In the following we will use $3+1$ metric variables to demonstrate the
existence of the constraints that remove the Boulware-Deser ghosts
from the physical spectrum.  This choice of variables is convenient
for isolating the nondynamical fields in the Einstein-Hilbert actions,
but their use requires some justification. In general, $\mathcal{N}$
Lorentzian metrics may not admit compatible notions of space and time
and compatible $3+1$ decompositions.  However, the vierbeine $e_I$ are
not fully independent since the parent theory gives rise to
constraints \eqref{symc}. These, as shown in \cite{Hassan:2017ugh},
insure that simultaneous $3+1$ decompositions exist for the pair $(u, 
e_I)$, for each $I$. Geometrically, the null cone of each $e_I$
shares some common timelike and spacelike directions with the null
cone of $u$ \cite{Hassan:2017ugh}. Therefore, with further mild
restrictions on the ranges of metric variables, there always exist
large classes of configurations where all $e_I$ share a common spatial
hypersuface with $u$, and hence, admit simultaneous $3+1$
decompositions. In the new multi-vierbein theory, the null cones are
further correlated by the equation \eqref{usol} for $u$.  The ghost
proof below assumes such configurations admitting simultaneous $3+1$
decompositions. It may be possible to further generalize it to a
covariant analysis on the lines of \cite{Deffayet:2012nr,
  Bernard:2015uic}. 

We decompose a vierbein $e_{~\mu}^{A}$ into a gauge-fixed vierbein
$E_{~\mu}^{A}$, rotated by a Lorentz transformation
$\Lambda^A_{~B}$,
\begin{align}\label{decomp}
e_{~\mu}^{A}=\Lambda_{~B}^{A}E_{~\mu}^{B}\,,
\end{align}
and use the following $3+1$ parameterization
\cite{Hinterbichler:2012cn, Hassan:2014gta, deRham:2015cha},
\begin{align}
E_{~\mu}^{A}&=
\begin{pmatrix}
  N ~&~ 0 \\
E_{~j}^{a}N^j~&~E_{~i}^{a}
\end{pmatrix}\,,
\nn\\
\Lambda^A_{~B}&=
\begin{pmatrix}
\gamma ~&~ v_c\\
v^a~&~\delta^{a}_{~c} +\frac{1}{1+\gamma}v^av_c
\end{pmatrix}
\begin{pmatrix}
1~&~0\\
0~&~({\Omega})_{~b}^{c}
\end{pmatrix}\,.
\end{align}
Here $N$ is the lapse, $N^i$ is the shift vector, and $E_{~i}^{a}$ is
a gauge fixed spatial vierbein with six independent components.  The
Lorentz transformation has been further decomposed into spatial
rotations parameterized by an $SO(3)$ matrix ${\Omega}$ (containing 
three parameters), and Lorentz boosts parameterized by the
3-dimensional (rescaled) boost vector $v^a$, with $\gamma\equiv
\sqrt{1+v_av^a}$.

In the action, we write $(\mathcal{N}-1)$ of the vierbeine
$(e_I)^A_{~\mu}$ using the $3+1$ paratemerization given above.  The
overall Lorentz invariance of the action allows us to take the last
vierbein to be of the same form but with $v^a=0$ and
${\Omega}=\mathbb{1}$.  Since the Lorentz parameters do not show up
in the kinetic terms, they appear without derivatives in the
action. Their equations are the Lorentz constraints
\eqref{constlor}.

It is easy to show that the potential \eqref{eff3} is linear in all
lapses $N_I$ and shifts $N^i_I$ before integrating out the Lorentz
fields $v_I^a$ and ${\Omega}_I$: The potential is $\det(u)$, where
the matrix $u$ is given by \eqref{usol}. The $N_I$ and $N^i_I$ appear
only in the first column $u^A_{~0}$. Since $\det(u)$ is antisymmetric
in the columns of $u$, each term in it contains only one factor of
$u^A_{~0}$, hence the linearity.  The Einstein-Hilbert terms are also
linear in these variables, hence, the entire action has the form,
\begin{align}\label{Sconst}
S=\sum_{I=1}^\mathcal{N}\int \dd^4 x\left(
(\pi_I)^i_{~a}(\dot{E}_I)^a_{~i}
- N_I \mathcal{C}_I-N_{Ii}\mathcal{C}^i_I
\right)\,.        
\end{align} 
Here, $(\pi_I)^{i}_{~a}$ are the canonical momenta conjugate to the
$(E_I)^a_{~i}$ and all other fields are nondynamical. The functions
$\mathcal{C}_I$ and $\mathcal{C}^i_I$ contain the $E_I$, $\pi_I$,
and the Lorentz fields, $v_I^a$ and ${\Omega}_I$, but not the 
lapses and shifts. We note that so far in this section the
considerations, including the form of the action in \eqref{Sconst},
are not restricted to the model \eqref{eff3}, but also apply to the
general class of multivielbein theories introduced in
\cite{Hinterbichler:2012cn}. However, the difference is crucial for
the argument that follows.

The dynamical fields $(E_I)^a_{~i}$ contain the ghost modes. These
must be eliminated by the constraints that arise from the equations of
motion for the $N_I$ and $N_{iI}$, that is, $\mathcal{C}_I=0$ and
$\mathcal{C}^i_I=0$, as well as from equations \eqref{constlor} for
the Lorentz fields, and the gauge fixing conditions for general
covariance. In particular it is necessary that, after solving for the
Lorentz fields, $\mathcal{C}_I=0$ become equations for the $E_I$ and
$\pi_I$ alone, remaining independent of the lapses. Then they can
eliminate the ghost fields in favour of the remaining dynamical
variables.  The difficulty is that for the general multivierbein
interactions of \cite{Hinterbichler:2012cn} which also have the form
\eqref{Sconst}, the Lorentz constraints
\eqref{constlor} render the spatial rotations ${\Omega}_I$, and
hence the $\mathcal{C}_I$, dependent on the $N_I$
\cite{deRham:2015cha}. Then, $\mathcal{C}_I=0$ can be solved for the
$N_I$ rather than eliminate the ghost fields which will remain in 
the spectrum. We now show that the multivierbein interaction in
\eqref{eff3} circumvents this problem.

First of all, $(\mathcal{N}-1)$ combinations of the ${\mathcal N}$
vector equations $\mathcal{C}^i_I=0$ can be used to determine the
$(\mathcal{N}-1)$ boost vectors $v_I^a$ as $v_I^a(E,\pi,{\Omega})$.
The unused combination of the $\mathcal{C}^i_I$, say $\mathcal{C}^i$,
are the three constraints of spatial diffeomorphisms. To determine the
rotation matrices ${\Omega}_I$, consider the Lorentz constraints
that for the model \eqref{eff3} give the $(\mathcal{N}-1)$ independent
matrix equations in (\ref{consteff}). The spatial parts of these are
the $3(\mathcal{N}-1)$ equations,
\begin{align}
\label{rotmatc}
\sum_{I=1}^\mathcal{N}(e_I)_{~[i}^{A}\eta_{AB}(e_J)_{~j]}^{B} =0\,,
~~~J=1,\hdots,\mathcal{N}-1\,.
\end{align}
Crucially, the $(e_I)_{~i}^{A}$ do not contain the lapses nor the
shifts since from (\ref{decomp}) it follows that,
\begin{align}
e_{~i}^{A}= 
\begin{pmatrix}
v_b{\Omega}^b_{~c}E_{~i}^{c}\\
\left(\delta_{~b}^{a}+\tfrac{1}{1+\gamma}v^av_b\right)
{\Omega}^b_{~c}E_{~i}^{c} 
\end{pmatrix}\,.
\end{align}
Therefore, the constraints (\ref{rotmatc}), along with the solutions
for $v_I^a$, determine the $3(\mathcal{N}-1)$ parameters of the
rotation matrices as ${\Omega}_I(E,\pi)$. This insures that 
$\mathcal{C}_I$ depend only on the $E_I$ and $\pi_I$, as desired. 

Finally, the $0i$ components of the Lorentz constraints provide
$3(\mathcal{N}-1)$ equations linear in the lapses and shifts. These
determine $(\mathcal{N}-1)$ of the shift vectors $N_I^i$ as linear
functions of the $N_I$ and of the ${\cal N}$th shift vector, say,
$N_{\cal N}^i$. The linearity insures that on using these solutions,
the action is linear in $N_I$ and $N_{\cal N}^i$.  At this stage, the
only other variables in the action are the $E_I$ and the $\pi_I$.

Of the remaining constraints, one combination of the $\mathcal{C}_I$,
say $\mathcal{C}$, together with the unused $\mathcal{C}^i$, form a
set of 4 first class constraints associated with general covariance,
as in \cite{Hassan:2018mbl}. The remaining ${\cal N}-1$ combinations
of the $\mathcal{C}_I$ are second class constraints. Their
preservation in time, $\dot{\mathcal{C}}_I=0$, gives another set of
${\cal N}-1$ constraints. Although we have not proven this here, their
existence can be argued in the parent theory (\ref{np1}) where the
calculations are very similar to the bimetric case explicitly analysed
in \cite{Hassan:2018mbl}.

The degree of freedom count is now easy. The fields $(E_I)_{~i}^{a}$
and the momenta $(\pi_I)^{i}_{~a}$ contain
$2\times6\mathcal{N}=12\mathcal{N}$ phase space variables, including
the ${\mathcal N}$ ghost fields and their ${\mathcal N}$ conjugate
momenta. The ${\cal N}-1$ second class constraints and their
associated time preservation conditions eliminate $2({\cal N}-1)$
ghost fields and ghost momenta. The 4 first class constraints and the
associated symmetry eliminate 8 phase space variables, including the
last pair of ghost variables, just as in general relativity. The
physical phase space thus consists of
$12\mathcal{N}-(2\mathcal{N}+6)=10\mathcal{N}-6$ variables,
corresponding to one massless field (with 4 phase space modes) and
$(\mathcal{N}-1)$ massive fields of spin-2 (each with $10$ phase space
modes).


\section{Existence of metric formulations}


Ghost-free multivierbein theories with only pairwise interactions can
be fully expressed in terms of the corresponding metrics. Here we
explore the existence of a metric formulation for the non pairwise
interaction in \eqref{eff3}. By extracting a factor of $\det e_1$,
and using $\gmn\equiv(e_1)_{~\mu}^{A}\eta_{AB}(e_1)_{~\nu}^{B}$, the
interaction becomes,
\begin{align}
\label{newform}
S_\mathrm{int}=-M^4 \int\dd^4x\sqrt{g}\,
\det\Big(\mathbb{1}+ g^{-1}  \sum_{J=2}^\mathcal{N}
e_1^\mathrm{T}\eta e_J\Big).
\end{align}
Defining the antisymmetric matrices $A_{IJ}$ as, 
\begin{align}
A_{IJ}\equiv \tfrac{1}{2}(e_I^\mathrm{T}\eta e_J-e_J^\mathrm{T}\eta e_I)\,,
\end{align}
the action depends on the linear combination $\sum_{J=1}A_{1J}$.
Also, in terms of these, the Lorentz constraints \eqref{consteff} read,
\begin{align}\label{lorca}
\sum_{J=1}^\mathcal{N}A_{IJ}=0\,,
\qquad I=1,\hdots,\mathcal{N}\,.
\end{align}
Hence, the antisymmetric parts drop out of (\ref{newform}). For any
pair of vierbeine, if we set $A_{IJ}=0$, then one could replace
$e_I^{-1} e_J= (g_I^{-1}g_J)^{-1/2}$, where $g_I=e_I^T\eta e_I$. Then
a covariant formulation in terms of the metrics would be possible if
$A_{1J}=0$ for all $J$. But the constraints \eqref{lorca} that arise
in the theory are weaker. Hence, although the $A_{IJ}$ drop out of the
action, the theory has no {\it equivalent} metric
formulation. However, metric formulations exist under mild
restrictions. 


\paragraph{Metric formulation for three fields.}
We now consider metric representations for the case $\mathcal{N}=3$,
following an analysis carried out in Ref.~\cite{Hassan:2012wt}.  Let
us introduce 3 Lorentz matrices $(\bar L_I)^{A}_{~B}$ as St\"uckelberg
fields for the 3 vierbeine $(e_I)_{~\mu}^A$, that is, we write every
vierbein as a Lorentz rotation of a gauge fixed vierbein
$(\bar{e}_I)_{~\mu}^A$,
\begin{align}
(e_I)_{~\mu}^A =(\bar L_I)_{~B}^{A}(\bar{e}_I)_{~\mu}^B\,. 
\end{align}
Defining $(L_I)_{~\mu}^{\nu}=(\bar{e}_I)^\nu_{~C}(\bar L_I)_{~B}^{C} 
(\bar{e}_I)_{~\mu}^{B}$, we can also write,
\begin{align}
(e_I)_{~\mu}^A=(\bar{e}_I)_{~\nu}^{A}(L_I)^{\nu}_{~\mu}\,.
\end{align}
Let us choose the gauge fixed vierbeine such that,
\begin{align}\label{sqrtcond}
(\bar{e}_I)_{~\mu}^A\eta_{AB}(\bar{e}_J)_{~\nu}^{B}=
(\bar{e}_I)_{~\nu}^{A}\eta_{AB}(\bar{e}_J)_{~\mu}^{B}\,,
\end{align}
for $I,J=1,2,3$.  General vierbeine may not be gauge fixed in this way
\cite{Deffayet:2012zc}, but this is possible for restricted field
configurations such that the null cones associated with the $I$ and
$J$ vierbeine intersect \cite{Hassan:2017ugh}. Note that this is a
stronger requirements than the existence of simultaneous $3+1$
decompositions assumed for the ghost analysis. Now, the Lorentz
constraints \eqref{consteff}, provide the following two independent
sets of equations for the $L_I$ which determine two of the three
Lorentz matrices,
\begin{align}\label{consteff2}
&L_1^T \bar{e}_1^\mathrm{T}\eta \bar{e}_2 L_2
-(1\leftrightarrow 2)
=L_3^T\bar{e}_3^\mathrm{T}\eta \bar{e}_1L_1
-(1\leftrightarrow 3)
\,,\nn\\
&L_2^T\bar{e}_2^\mathrm{T}\eta \bar{e}_3 L_3
-(2\leftrightarrow 3)
=L_3^T\bar{e}_3^\mathrm{T}\eta \bar{e}_1L_1
-(1\leftrightarrow 3)\,.
\end{align} 
Multiplying the first with $L_1^{-1T}$ from the left and $L_1^{-1}$
from the right and the second with $L_3^{-1T}$ from the left and
$L_3^{-1}$ from the right, it is obvious that the covariant solution
is $L_1L_{2}^{-1}=L_1L_{3}^{-1}=L_3L_{2}^{-1}=\mathbb{1}$ which
implies $L_1=L_2=L_3$. Then the multi-spin-2 potential becomes,
\begin{align}
V=M^4 \det \left(\bar{e}_1 +\bar{e}_2 +\bar{e}_3 \right)\,, 
\end{align}
since the undetermined overall Lorentz matrix drops out due to 
Lorentz invariance. We can extract a factor of $\det e_1$ to
obtain, 
\begin{align}\label{e1act}
  V=M^4\det(e_1)\det\left(\mathbb{1}+\bar{e}_1^{-1}\bar{e}_2
  +\bar{e}_1^{-1}\bar{e}_3\right)\,.
\end{align}
Let us introduce the metrics $\gmn$, $\fmn$, and $\hmn$, for the
vierbeine $\bar{e}_1$, $\bar{e}_2$, and $\bar{e}_3$, respectively. Due
to the symmetrization constraint (\ref{sqrtcond}), we have that,
\begin{align}
\bar{e}_1^{-1}\bar{e}_2=\sqrt{g^{-1}f}\,,\qquad
\bar{e}_1^{-1}\bar{e}_3=\sqrt{g^{-1}h}\,,
\end{align}
Then the multi-spin-2 potential can be written in terms of metrics as,  
\begin{align}
\label{fineffpot}
V=M^4\sqrt{g}\det\left(\mathbb{1}+\sqrt{g^{-1}f}+
\sqrt{g^{-1}h}\right)\,. 
\end{align}
We could have chosen to extract the determinant of a vierbein
other than $e_1$ and correspondingly obtain,
\begin{align}
  V &= M^4 \sqrt{f}\det\left(\mathbb{1}+\sqrt{f^{-1}g}+
  \sqrt{f^{-1}h}\right) \nn\\
    &= M^4 \sqrt{h}\det\left(\mathbb{1}+\sqrt{h^{-1}g}+
  \sqrt{h^{-1}f}\right)\,. 
\end{align}
The above considerations partially generalize to $\mathcal{N}>3$.
Now the gauge choices \eqref{sqrtcond} cannot be made for all
vierbeine since these are $\mathcal{N}(\mathcal{N}-1)/2$ conditions,
while there are at most $\mathcal{N}$ Lorentz transformations to
implement the symmetrizations. The remaining conditions may be imposed
by hand on the $\bar{e}_I$, but this is not necessary.  To find a
metric formulation similar to, say, \eqref{fineffpot}, it is enough to
impose \eqref{sqrtcond} only on the
$(\bar{e}_1)_{~\mu}^A\eta_{AB}(\bar{e}_J)_{~\nu}^{B}$ which gives
$\mathcal{N}-1$ gauge conditions. Then, with appropriate restrictions,
and on choosing a similar solution for the $L_I$, one obtains an
expression which is a direct generalization of \eqref{fineffpot} to
${\mathcal N}$ metrics.

Finally, it is important to note that while the vierbeine were
restricted by hand to obtain a metric formulation, such restrictions
are inbuilt in the final multimetric theory. Hence, the resulting
multimetric theories can be considered in their own right, independent 
of the starting vierbein formulations.


\section{Discussion}


To summarize, we have constructed nontrivial interactions of multiple
spin-2 fields, beyond the known pairwise potentials, and have
demonstrated the existence of constraints that eliminate the extra
ghost modes. The interactions are given in terms of the ${\mathcal
  N}$ vierbeine $(e_I)^A_{~\mu}$ in equation \eqref{vbint}. On
expressing the determinant in terms of the wedge products of the
one-forms $e_I^A$, one gets,
\begin{align}
\label{wedge}
\sum_{I,J,K,L=1}^\mathcal{N} \beta^{IJKL} ~
\epsilon_{ABCD}~e_I^A \wedge e_J^B \wedge e_K^C \wedge e_L^D\,,
\end{align}
with $\beta^{IJKL}=\beta^I\beta^J\beta^K\beta^L $. Such interactions
with general $\beta^{IJKL}$ were proposed in
\cite{Hinterbichler:2012cn} where the $\mathcal{N}=2$ case reproduces
the bimetric theory \cite{Hassan:2011zd}. However, the only ghost-free
cases known so far were the pairwise bimetric interactions
\cite{deRham:2015cha}.  Given the interactions in \eqref{wedge} with
arbitrary $\beta^{IJKL}$, a direct analysis that identifies the
ghost-free cases is so far not known. But the construction presented
here generates a class of genuinely multivierbein ghost-free
interactions with up to 4 different vierbeine in one vertex. This
trivially extends to $D$ space-time dimensions, in which case the
vertices would contain up to $D$ different fields.

The class of theories obtained here can be further generalized.
First, note that the interactions \eqref{vbint} cannot be simply added
to a theory involving the previously known ghost-free interactions of
the $\mathcal{N}$ vierbeine $(e_I)_{~\mu}^{a}$.  Such a setup would
correspond to forbidden loop couplings in the parent theory with
vierbein $u_{~\mu}^{a}$.  However, compatible interactions can be
constructed by extending the parent theory by additional allowed
pairwise couplings as will be discussed in \cite{toappear}. The new
terms couple any of the vierbeine $e_I$ to an additional set of
$\mathcal{N}'$ vierbeine $v_K$ and include the following.
\vspace{-.2cm}
\begin{enumerate}[leftmargin=*]
\item One of the $e_I$ can interact with the $v_K$ through the
  standard pairwise interactions, that is the potential can have the
  form $V\sim\det(\sum_Ie_I)+\sum_KV_\mathrm{bi}(e_1, v_K)$.
\vspace{-.7cm}
\item One of the $e_I$ can interact with the $v_K$ through a
  determinant interaction, {\it i.e}., the potential could be
  $V\sim\det(\sum_Ie_I)+\det (e_1+\sum_Kv_K)$. 
\end{enumerate}
\vspace{-.2cm}

It is straightforward to introduce a standard coupling to matter via
any of the dynamical vierbeine $(e_I)_{~\mu}^{a}$ as this will not
influence the procedure of integrating out the non-dynamical vierbein
$u_{~\mu}^{a}$. On the other hand, if $u_{~\mu}^{a}$ couples to
matter, then the multi-spin-2 theory for the $(e_I)_{~\mu}^{a}$ will
have matter interactions that are heavily modified
\cite{Luben:2018kll}. At low energies these reduce to the matter
coupling suggested in \cite{Noller:2014sta, Hinterbichler:2015yaa}.

The interactions for $(\mathcal{N}+1)$ vierbeine that we started with
in \eqref{intact} were not of the most general ghost-free form. It
would be interesting to extend our setup to more general interactions
and obtain possibly ghost-free multi-spin-2 theories that are
different from the one studied here. It is not obvious that the
algebraic equations for the non-dynamical vierbein can be solved
covariantly for more general parameter choices in the action. In any
case, if the linear relation in \eqref{usol} is lost, the resulting
interactions may not have the general form in \eqref{wedge}. It is
important to find out if such ghost-free multivierbein interactions,
beyond the classes discussed here, could exist.

Our couplings are the first instance of ghost-free spin-2 interactions
where the vierbein formulation admits more general configurations than
the associated, more restrictive, metric formulation.  Hence it is
interesting to directly investigate the associated multimetric
theories, without recourse to the vierbein formulation. One expects
that the extra restrictions on the metrics, already encoded in the
multimetric interactions \cite{Hassan:2017ugh}, would lead to better
causal properties for the theory.



\end{document}